\documentclass[sigconf]{acmart}

\renewcommand\footnotetextcopyrightpermission[1]{} 
\usepackage[]{graphicx}
\usepackage[]{color}
\makeatletter
\def\maxwidth{ %
 \ifdim\Gin@nat@width>\linewidth
 \linewidth
 \else
 \Gin@nat@width
 \fi
}
\makeatother

\usepackage{subfig}
\usepackage{listings}
\makeatletter
 {\par\unskip\endMakeFramed%
 \at@end@of@kframe}
\makeatother

\usepackage{alltt}
\usepackage{multicol}
\AtBeginDocument{%
 \providecommand\BibTeX{{%
 \normalfont B\kern-0.5em{\scshape i\kern-0.25em b}\kern-0.8em\TeX}}}

\setcopyright{none}

\acmConference[Presented at DYNAMICS '19]{DYNAMICS '19: DYnamic and Novel Advances in Machine Learning and Intelligent Cyber Security Workshop}{December 09--10, 2019}{San Juan, PR}

\newcommand\blfootnote[1]{%
  \begingroup
  \renewcommand\thefootnote{}\footnote{#1}%
  \addtocounter{footnote}{-1}%
  \endgroup
}

\begin{document}

\title{Traffic Generation using Containerization for Machine Learning}

\author{Henry Clausen}
\email{henry.clausen@ed.ac.uk}
\affiliation{%
  \institution{University of Edinburgh}
    \city{Edinburgh, UK}
}

\author{Robert Flood}
\email{s1784464@ed.ac.uk}
\affiliation{%
  \institution{University of Edinburgh}
      \city{Edinburgh, UK}
}

\author{David Aspinall}
\affiliation{%
  \institution{University of Edinburgh}
      \city{Edinburgh, UK}
  \\ \institution{The Alan Turing Institute}
      \city{London, UK}
}

\begin{abstract}

The design and evaluation of data-driven network intrusion detection methods are currently held back by a lack of adequate data, both in terms of benign and attack traffic. Existing datasets are mostly gathered in isolated lab environments containing virtual machines, to both offer more control over the computer interactions and prevent any malicious code from escaping. This procedure however leads to datasets that lack four core properties: heterogeneity, ground truth traffic labels, large data size, and contemporary content.
Here, we present a novel data generation framework based on Docker containers that addresses these problems systematically. For this, we arrange suitable containers into relevant traffic communication scenarios and subscenarios, which are subject to appropriate input randomization as well as WAN emulation. By relying on process isolation through containerization, we can match traffic events with individual processes, and achieve scalability and modularity of individual traffic scenarios.
We perform two experiments to assess the reproducability and traffic properties of our framework, and demonstrate the usefulness of our framework on a traffic classification example.
\end{abstract}

\keywords{Network security, datasets, machine learning, intrusion detection}

\maketitle

\section{Introduction}

\blfootnote{\small This work was presented at the \textbf{ACSAC DYNAMICS '19}: DYnamic and Novel Advances in Machine Learning and Intelligent Cyber Security Workshop in December 09-10, 2019, San Juan, PR, and will be published in the corresponding workshop proceedings. This document version is specifically for publication on \textbf{arXiv}.}
The exponential growth of data availability enabled the machine learning revolution of this decade that transformed many areas of our lives. Ironically, security oriented data describing computer networks is notoriously hard to obtain, and researchers struggle to evaluate new network intrusion detection systems (NIDS) or similar tools on suitable network traffic data.
Well-designed datasets are such a rarity that researchers often rely on datasets that are well over a decade old \cite{tavallaee2009detailed, kayacik2005selecting}, calling into question their effectiveness on modern traffic and attacks. The lack of quantity, variability, meaningful labels, and ground truth has so far slowed scientific progress and objective and appropriate measurements on ML-based network security methods.

Privacy and security concerns discourage network administrators to release rich and realistic datasets for the public. Network traffic produced by individuals contains a mass of sensitive, personal information, such as passwords, email addresses, or usage habits, requiring researchers to expend effort anonymising the dataset \cite{mirsky2016sherlock}. To examine malicious behaviour, researchers are often forced to build artificial datasets using isolated machines in a laboratory setting to avoid damaging operational devices. Background traffic is usually generated in real-time from scripts executed on the virtual machine, which constrains both the amount and heterogeneity of the data. Current experiments show that it is possible to use large network of virtual machines for traffic generation \cite{crussell2019lessons,crussell2019virtually}, however we did not find any publicly available dataset of that category. 

Existing network intrusion detection datasets are predominantly designed to support a broad range of applications, and are collected in a static manner, unable to be modified or expanded. This proves to be a serious defect as the ecosystem of intrusions is continually evolving.
Furthermore, it prohibits a more detailed analysis of specific areas of network traffic due to the available data only being a fraction of the original dataset. To combat this, new datasets must be periodically built from scratch.

Allowing researchers to create datasets dynamically to circumvent these issues would be extremely beneficial. We propose a such framework, based on application containers using Docker \cite{docker}. Docker is a service for developing and monitoring containers, also known as OS-level virtual machines. 
By moving from virtual machines to containers, we enable the scalable, modular, and dynamic creation of network traffic datasets. Since Docker containers can be arranged in complex settings with a few commands, it is a lot easier with containers to script a variety of network activities thus increase the heterogeneity and realism of the generated data.

Furthermore, each Docker container is highly specialized in its purpose, generating traffic related to only a single application process. Therefore, by scripting a variety of Docker-based \emph{scenarios} that simulate benign or malicious behaviours and collecting the resultant traffic, we can build a dataset with perfect ground truth, something that has so far not been possible for network traffic. 

Finally, the most import reason to rely on Docker for containerization is that many containerized applications are shared on the Docker Hub platform.


This work provides the following contributions:

\begin{enumerate}
 \item We present a novel network traffic generation framework that is designed to improve several shortcomings of current datasets for NIDS evaluation. This framework is openly accessible for researchers and allows for straightforward customization.
 \item We define four new requirements a network intrusion dataset should fulfil in order to be suitable to train machine-learning based intrusion detection methods. 
 \item We perform a number of experiments to demonstrate the suitability and utility of our framework. 
\end{enumerate}

\subsection{Outline}

The remainder of the paper is organized as follows. Section \ref{Sec:background} discusses existing NIDS datasets and the problems that arise during their usage as well as background information about network traffic data formats and virtualization methods. The section concludes with a set of requirements we propose to improve the training and evaluation of machine-learning-based methods. Section \ref{Sec:Design} describes the general design of our framework, and how it improves on the discussed problems in existing datasets. We also discuss a specific example in detail. Section \ref{Sec:Experiments} discusses several experiments to validate the improvements and utility our framework provides. 
Section \ref{Sec:Conclusion} concludes the results and discusses limitations of our work and directions for future work.

\section{Background}\label{Sec:background}

\subsection{Data formats}

Computers in a network communicate by sending \emph{network packets} to each other, which are split into the control information, also called packet header, and the user information, called payload. The payload of a packet in general carries the information on behalf of an application and can be encrypted, while the header contains the necessary information for the correct transmission of the packet, including the transmission protocol layer, IP addresses, etc. Methods using packet-level input can be divided into payload inspection, header-based, or hybrid. Packets are usually stored in the widespread \emph{pcap} format.

The majority of packets are exchanged between two hosts within bidirectional \emph{connections}. Another common format of network traffic information is based on connection summaries, also called \emph{network flows}. RFC 3697 \cite{brownlee1999traffic} defines a network flow as a sequence of packets that share the same source and destination IP address, IP protocol, and for TCP and UDP connections the same source and destination port. A network flow is usually represented by this information along with additional information such as the start and duration of the connection as well as the total number of packets and bytes transferred. 



\subsection{Related work and existing datasets}

To evaluate their ability to model the behaviour of a network and to identify malicious activity and network intrusions, new methods have to be tested using existing datasets of network traffic. This network should ideally contain realistic and representative benign network traffic as well as a variety of different network intrusions. However, as network traffic contains a vast amount of information about a network and its users, it is notoriously difficult to release a comprehensive dataset without infringing the privacy rights of the network users \cite{sperotto2009labeled}. Furthermore, the identification of malicious traffic in network traces is not straightforward and often requires a significant amount of manual labelling work. Introducing malicious software into an enterprise network would be impossible for ethical and security reasons. For that reason, only about four structured datasets for network intrusion containing real-world traffic and attacks are available openly. The most recent and notable real-world datasets have been released from the Los Alamos National Laboratory (LANL) in 2015 and 2017 \cite{akent-2015-enterprise-data, turcotte17}, and the University of Granada (UGR) in 2016 \cite{macia2018ugr}. Both datasets contain network flow traffic data from a large number of hosts collected over multiple months, giving an accurate representation of medium- to large-scale structures in benign traffic. However, the amount of attack data is small and insufficient for accurate detection rate estimation. Furthermore, packet-level data is not available for both datasets. Other real-world datasets, such as CAIDA 2016 \cite{walsworth2015caida} or MAWI 2000 \cite{sony2000traffic}, provide packet headers, but are unstructured and contain no labeled attack data at all.

To improve the lack of attack traffic in NIDS datasets, several artificially created datasets have been proposed. For this, a testbed of virtual machines is usually hosted in an enclosed environment to prevent any malicious code from spreading to other machines on other networks. To generate attack traffic, these machines are then subject to a selection of attack carried out by other machines in the environment. Benign traffic is generated using commercial traffic generators such as the \emph{IXIA PerfectStorm tool}, or by scripting a selection of tasks for each machine. Synthetic datasets cover a smaller timeframe and contain traffic from a small number of hosts. Notable examples are the CIC-IDS 2017 dataset from the Canadian Institute for Cybersecurity \cite{sharafaldin2018towards}, 
and the UNSW-NB 2015 dataset from the University of New South Wales \cite{moustafa_unsw-nb15:_2015}. Both datasets contain traffic from a variety of attacks, and are available as packet headers or as network flows with additional features crafted for machine-learning. While the benign traffic for the CIC-IDS 2017 data was generated using scripted tasks from a number of host profiles, the benign data for the UNSW-NB 2015 data is a mixture of captured real traffic from another subnet and traffic generated using a commercial traffic replicator.

We omitted the synthetic KDD-Cup 1999 and the DARPA 1998 datasets along with their derivates from the discussion as they are well-known to be outdated and contain unrealistic benign traffic, artificially high benign/attack data ratios, and artifacts stemming from communication simulations \cite{tavallaee2009detailed,mchugh2000testing}, problems which have been addressed by most modern datasets.

Container networks have recently been adopted to conduct traffic generation experiments, such by Fujdiak et al. \cite{fujdiak2018ip} who use containerized web servers to collect DoS-traffic. Furthermore, significant effort has been put into the creation of large-scale virtualization frameworks to provide automatized network testbeds \cite{crussell2015minimega, badiger2018violet}.

\subsection{Problems in modern datasets}\label{Sec:problems}

The difficulty of obtaining malicious traffic in real-world captures means that the performance of new network intrusion detection algorithms are almost exclusively evaluated on synthetic datasets. Potential disadvantages of synthetic compared to real-world datasets have been discussed by several authors \cite{sommer2010outside,sperotto2009labeled}. However, none address problems in the particular design of such synthetic testbeds that are holding machine-learning based methods back in performance and from getting more widespread application. Here, we focused on this aspect and four design problems common among modern synthetic datasets.

\paragraph{Lack of variation}\label{Sec:lackvar}

To generate benign traffic, a selection of activities is scripted and executed on virtual machines. Activities are selected to cover the most prominent protocols, but seldom to cover the range of subactivites that each protocol offers. Instead, the manner in which each protocol is used is highly-restricted, and there are doubts about whether this traffic is representative of its real-world equivalent usage \cite{sommer2010outside}. 
An illustrative example of the restricted protocol activity in synthetic datasets can be seen in the CIC-IDS 2017 dataset. Here, the vast majority of successful FTP transfers consist of a client downloading a single text file containing the Wikipedia page for `Encryption' several hundred times in a day. In reality, FTP is used for a large number of tasks, which can occur in random order with varying input sizes and parameters. 

In addition to that, implemented test bed environments are usually separated from external influence or even virtualized, which isolates them from fluctuations and faults introduced by the complexity of modern networking. These include packet delays through network congestions, unexpected connection drops or resets, and out-of-order arrivals, all of which lead to variations in the response behavior of particular services.

This general lack of variation in individual protocols leads to observed homogeneity both on a packet exchange level and on a network flow level, and thus to clearer structures in the data. Identifying separations of malicious and benign activity or between different services consequently becomes easier, which leads to overoptimistic results in the evaluation of machine-learning based methods. It is therefore clear that traffic variation is a crucial aspect of a comprehensive intrusion detection dataset.
parameter

\paragraph{Lack of ground truth}

To evaluate machine-learning-based methods that distinguish between different types of network traffic data, we need to verify that separating structures in the model correspond to distinct computational actions, using traffic labels. Most obvious is the labelling of benign and malicious traffic. More granular labels are desirable to distinguish between several different types of network traffic. An example for this is the design of `stepping stone' detection methods, where researchers try to detect connections relayed over a jump-host. Similarity or correlation metrics that measure the closeness of two connections are a popular tool. To understand such a measure, ground truth about how computationally similar two connections are and what type of behavior they represent is necessary. 
Other areas that look at small-scale traffic structures and would benefit from detailed traffic labels include protocol verification, traffic classification, traffic disaggregation, or exploit discovery.

Ground truth labels for network traffic are hard to obtain. The network traffic produced by a typical PC will invariably contain traffic originating from background processes, such as software updates, authentication traffic, network discovery services, advertising features, as well as many other sources. To separate traffic from different origins retrospectively is often hard, if not impossible. Source attribution through port numbers is unreliable because port numbers can be dynamically allocated and are not restricted to particular processes, and processes can open connections on multiple ports at the same time. 
All of these reasons mean that the identification of different computational operations from captured traffic is often infeasible. 
Therefore, no public NID dataset currently considers the inclusion of ground truth traffic labels.

\paragraph{Static design}

A released dataset can only contain data that is representative a system at the time of creation. In contrast to other many other data sources for machine-learning, network traffic, both benign and malicious, is constantly changing as computational protocols and systems evolve. All available NIDS datasets today have been created in a static manner , so that a fixed test bed of host machines is created designed to contain specific vulnerabilities to the selected attacks. This makes it very hard to change the test bed and thus adjust the dataset to updated traffic structures. Allix et al. \cite{allix2014machine} claim that it is impossible to release a NID dataset that is truly representative of the real-world attacks due to the inherent secrecy of the intrusion ecosystem and the rate at which it develops.

\paragraph{Limited size}
Today's machine-learning revolution was supercharged by the exponential growth of available data. Larger amounts of data mean that a given model can identify more complex structures that remain invariant in noisy environments and thus generalize better. Although the amount of globally transmitted network traffic is growing every year, the size of available NIDS datasets is limited by small host numbers, typically 5-10, and short capture periods, at maximum a 5-6 weeks, inherent to test bed captures. This means that traffic models can experience difficulties to generalize over specific traffic types which represent a smaller fraction of the total dataset. In an ideal setting, researchers would have the ability to generate arbitrary amounts of specific traffic types.

\subsection{Containerization with Docker}
Virtual machines (VMs) share the same hardware infrastructure as the host machine. VMs necessitate the use of hypervisors, software responsible for sharing the host OS's hardware resources, such as memory, storage and networking capabilities. OS-level virtualization, also known as \textit{containerization}, is a virtualization paradigm that has become popular in recent years due to its lightweight nature and speed of deployment. 
In contrast with standard VMs, containers forego a hypervisor and the shared resources are instead kernel artifacts, which can be shared simultaneously across several containers. Although this prevents the host environment from running different operating systems, containerization incurs minimal CPU, memory, and networking overhead whilst maintaining a great deal of isolation \cite{kolyshkin2006virtualization}.

The main advantage of using containers for traffic generation is the isolation of individual applications. This enables us to gather ground truth about the traffic origin, and enables us to easily extend, modify, and scale our traffic generation framework, which would not be possible when relying on VMs.


\paragraph*{Docker container}

\textit{Docker} is a software platform that allows for the creation, maintenance and deployment of containers. In Docker's terminology, a container is a single, running instance of a Docker \textit{image}. Docker images are defined via a text file known as the \texttt{Dockerfile}, which consists a series of commands that modify an underlying \textit{base image}, usually a containerized OS. Example commands include installing libraries and copying files. Figure \ref{fig:dockerfile} displays a simple example of \texttt{Dockerfile}.

\begin{figure}
\begin{lstlisting}[frame=tb,frame=single,caption=,label=code1,basicstyle=\footnotesize]
FROM ubuntu
MAINTAINER XYZ (email@domain.com)
RUN apt-get update
RUN apt-get install -y nginx
ENTRYPOINT ["/usr/sbin/nginx","-g","daemon off;"]
EXPOSE 80
\end{lstlisting}
 \caption{Example of \texttt{Dockerfile} creating a nginx-container.}
 \label{fig:dockerfile}
\end{figure}

After each command is executed, the intermediate, read-only image is saved as a \textit{layer}. These layers can be shared between containers. When a Docker image is run as a container, a final read-write layer is added and when the container is later stopped, this layer is discarded, preserving the integrity of the underlying layers. This allows Docker containers to be run repeatedly whilst always starting from an identical state.

Individual Docker containers are intended to be highly specialized in their purpose with each container running only a specific piece of software or application. Commonly used base images --- such as Alpine Linux --- have minimal background processes running during a container's lifetime. This means that the network traces of a Docker container can be associated with a specific application. The one-to-one correlation between containers and network traces allows us to produce labeled datasets with fully granular ground truths.

The Docker software platform includes a cloud-based repository called the Docker Hub \cite{dockerhub} which allows users to download and build open source images on their local computers. At the time of writing, nearly 2.5 million images are available from the Docker Hub. Some common software --- such as popular webservers and databases --- have officially maintained images. We use these as far as possible to simplify the production of our scenarios, and keep them close to software configuration used in practice.


\paragraph*{Docker Networking} 
\label{sec:network}
Docker allows the creation of virtualized networks with one or more subnetworks, to which containers can connect via a virtualized network bridge. Containers attached to the bridge network are assigned an IP address and are able to communicate with other containers on their subnetwork. Containers can furthermore be connected to a host network, which allows communication with external networks using NAT via the host interface.

To host containers in an isolated network, we can create our own user-defined bridge networks, which provides greater isolation between containers \cite{docker_docs}. Furthermore, this allows us to fix the subnet and gateway for our networks as well as the IP addresses of our containers, which simplifies scripting scenarios. Docker allows containers to share the same network interface. This enables us to assign the same IP address to multiple containers.

\paragraph*{Netem} 
The Docker engine also provides network access to Linux traffic control facilities such as \emph{NetEm}, on which we will rely in this project. NetEm is a Linux toolkit for testing protocols by emulating properties of wide area networks \cite{hemminger2005network}. It allows the user emulates variable delay, loss, duplication and re-ordering of packets on particular network interfaces.

\paragraph*{Docker Compose}
Applications built using the Docker framework often need more than one container to operate, for example an Apache server and a MySQL server running in separate containers. We must build and deploy several interconnected containers simultaneously. Docker provides this functionality via \texttt{Docker compose}, a tool that allows users to define the services of multiple containers as well as the properties of virtual networks in a YAML file. By default, this file is named \textit{docker-compose.yml}. This allows for numerous containers to be started, stopped and rebuilt with a single command in a consistent manner. This is particularly significant for our purposes; with \texttt{Docker compose}, we can launch several containers in a specific order, with a specific network configuration, whilst running specific commands within each container on start up. This ensures that our interactions are deterministic, barring any added randomization. Figure \ref{fig:dockercompose} displays an example of a simple \texttt{Docker compose} file.

\begin{figure}
\begin{lstlisting}[frame=tb,frame=single,caption=,label=code2,basicstyle=\footnotesize]
version: '3'
services:
 webserver:
 image: nginx:alpine
 ports:
 - "80:80"
 networks:
 - app-network
 db:
 image: mysql:5.7.22
 ports:
 - "3306:3306"
 environment:
 MYSQL_DATABASE: laravel
 MYSQL_ROOT_PASSWORD: your_mysql_root_password
 networks:
 - app-network
networks:
 app-network:
 driver: bridge
\end{lstlisting}
\caption{Example of \texttt{Docker-compose} file launching a nginx- and a mysql-container in an isolated network.}
\label{fig:dockercompose}
\end{figure}{}

\subsection{Dataset Requirements}\label{Sec:require}

The primary task of this project is to provide a suite of Docker container compositions that is capable of generating traffic datasets suitable for machine-learning-based intrusion detection systems. This container suite is designed to address the criticism of current NIDS datasets discussed in Section \ref{Sec:problems}. For this, we created a set of requirements that a modern intrusion detection dataset has to fulfil to address the problems discussed in Section \ref{Sec:problems}:

 \paragraph{Variation} To ensure that we produce representative data for modelling, we want the traffic generated by our container suite to cover a sufficient number of protocols that are commonly found in real-world traffic and existing datasets. For malicious traffic, we want to ensure that the attacks are modern and varied, both in purpose and in network footprint. For each protocol, we want to establish several capture scenarios to encompass the breadth of that protocol's possible network traces. Communication between containers should be subject to the same disturbances and delays as in a real-world setting.


\paragraph{Ground truth} Since ground truth is a main focus of this work, we want a capture scenarios to be consistent and reproducible in the traffic they generate. This way, we can be certain that a particular traffic trace corresponds to the capture scenario it was generated by, and can thus relate individual traffic events to computational operations. We discuss what it means for a scenario to be reproducible in detail in Section \ref{Sec:deterministic}. 

\paragraph{Modularity} Traffic capture scenarios should be implemented in a modular way allow for a straightforward addition or modification of traffic capture modules without disrupting the rest of the container suite. This reduces the effort to adjust a dataset to changing traffic patterns and allows the addition of modern attacks traffic.

\paragraph{Scalability} Each capture scenario should be running in a scalable manner to allow generation of large data quantities.

\section{Design}\label{Sec:Design}

To cover a range of activities, the containers in our framework are arranged in different configurations corresponding to particular \emph{capture scenarios}. Running a given capture scenario triggers the launch of several Docker containers, each with a scripted task specific to that capture scenario. A simple exemplary capture scenario may consist of a containerized client pinging a containerized server. We ensure that each Docker container involved in producing or receiving traffic will be partnered with a \texttt{tcpdump} container, allowing us to collect the resulting network traffic from each container's perspective automatically. 

\begin{figure}
 \centering 
 \includegraphics[width=0.480\textwidth]{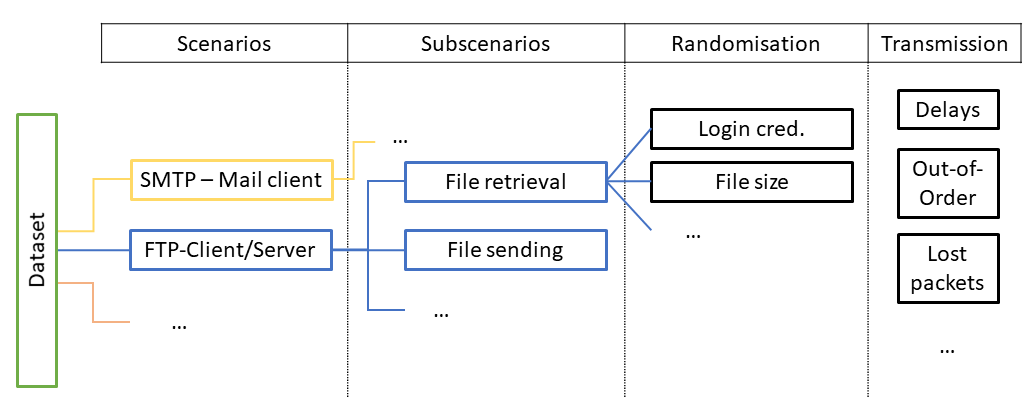}
 \caption{Visualization of the different levels at which traffic variation is introduced in DetGen.}
 \label{Fig:branching}
\end{figure}

We outline different stages within the creation of a dataset at which traffic variation is introduced. Figure \ref{Fig:branching} visualizes this process.

\subsection{Scenarios}
\label{Sec:Scenarios}

We define a \emph{scenario} as a series of Docker containers interacting with one another whereby all resulting network traffic is captured from each container's perspective. This constructs network datasets with total interaction capture, as described by Shiravi et al. \cite{shiravi2012toward}. Each scenario produces traffic from either a protocol, application or a series thereof. Both benign and malicious activities are implemented as scenarios. Examples may include an FTP interaction, a music streaming application and client, an online login form paired with an SQL database, or a C\&C server communicating with an open backdoor. A full list of currently implemented scenarios can be found in Section \ref{Sec:ExistScen}.

Each scenario is designed to be easily started via a single script that allows the user to set the length of the capture time, and the specification of particular subscenarios, discussed below. Scenarios can be repeated indefinitely without further instructions and be run in parallel, therefore allowing the generation of large amounts of data.

Our framework is modular, so that individual scenarios are configured, stored, and launched independently. Adding or reconfiguring a scenario has no effect on the remaining framework.

\subsection{Subscenarios} \label{Sec:Subscenarios}

In contrast to scenarios, \textit{subscenarios} provide a finer grain of control over the traffic to be generated, allowing the user to specify the manner in which a scenario should develop. The aim of having multiple subscenarios for each scenario is to explore the full breadth of a protocol or application's possible traffic behavior. For instance, the SSH protocol can be used to access the servers console, to retrieve or send files, or for port forwarding, all of which may or may not be successful. It is therefore appropriate to script multiple subscenarios that cover this range of tasks.

The same applies to malicious activity. For instance, it would be naive for an SSH password bruteforcing scenario to always successfully guess a user's password. Instead, we include a second subscenario in which the password bruteforcer fails.

Subscenarios are specific to particular scenarios and can be specified when launching that scenario.

\subsection{Randomization within Subscenarios}\label{Sec:randomsubscen}

Scripting activities that are otherwise conducted by human operators often leads to a loss of random variation that is normally inherent to the activity.
As mentioned in Section \ref{Sec:problems}, the majority of successful FTP transfers in the CIC-IDS 2017 data consist of a client downloading a single text file. In reality, file sizes, log-in credentials, and many other variables included in an activity are more or less drawn randomly, which naturally influences traffic quantities such as packet sizes or numbers.

To account for these fluctuations, we identify variable input parameters within scenarios and their subscenarios and systematically draw them randomly from a suitable distribution. Passwords and usernames, for instance, are generated as a random sequence of letters with a length drawn from a Cauchy distribution, before they are passed to the corresponding container. Files to be transmitted are selected at random from a larger set of files, covering different sizes and file names.

\subsection{Network transmission}\label{Sec:Netrand}

 Docker communication takes place over virtual bridge networks, so the throughput is far higher and more reliable than in real-world networks, with the Docker virtual network achieving a bandwidth of over 90 Gbits/s when measured using iPerf \cite{iperf}. This level of speed and consistency is worrying for our purposes as packet timings will be largely identical on repeated runs of a scenario and any collected data could be overly homogeneous.

To retard the quality of the Docker network to realistic levels, we rely on emulation tools. As discussed in section \ref{sec:network}, Netem is a Linux command line tool that allows users to artificially simulate network conditions such as high latency, low bandwidth or packet corruption in a flexible manner.

Although it is relatively straightforward to apply Netem commands to a Docker Bridge network, we decided not to invoke Netem in this manner as this would cause all network settings of all containers to be identical, such as all containers in a scenario having a latency of 50ms.  Instead, we developed a wrapping script that applies Netem commands to the network interface of a given container, providing us with the flexibility to set each container's network settings uniquely. This script randomizes the values of each parameter, such as packet drop rate, bandwidth limit, latency, ensuring that every run of a scenario has some degree of network randomization if desired.

\subsection{Capture}


To capture traffic, we use containers running \texttt{tcpdump}, a widespread and free packet analyzer software that can capture packets arriving at or leaving from a network interface \cite{jacobson1989tcpdump}.
We attach \texttt{tcpdump} containers on every interface in the virtualized docker network and write packets into separate capture files. This allows us to capture traffic from the perspective of every container in a scenario, giving a complete view. 
In Section \ref{Sec:datasetcreation}, we discuss how the collected capture files are coalesced into one dataset.


\subsection{Implementation Process}

The implementation process for each scenario follows broadly the same outline: 

\begin{enumerate}

\item Select containers which provide the required services and identify the \textit{primary} container/s for a given scenario which is/are dictating the container interaction. Then create and build a Dockerfile containing all necessary dependencies. 

 

\item Identify different ways to use the service of the given scenario and define them into a set of subscenarios. 

\item Design and implement the behavior for secondary containers to provide the required service to the primary container(s).

\item For each subscenario, identify variable input values and their appropriate range; then systematically implement their generation from appropriate distributions covering this range.
 
\item Add \texttt{tcpdump} containers to every network interface.

\item Create a \texttt{Docker compose} file that launches all containers simultaneously.

\item Finally, write a script that, upon running, calls this \texttt{Docker compose} file, applies a network emulation script to each container network interface, and allows the user to specify how long and how many times a scenario should be run. 

\end{enumerate}

Following the Docker guidelines \cite{bestpractise}, each container in our framework consists of a single service with a specialized purpose, with as few additional dependencies as possible. 
Moreover, we ensure that there are minimal inter-dependencies between the containers of a scenario. This allows us to easily modify and update containers as new versions of the underlying software are released.

\subsection{Simple Example Scenario - FTP server}
\begin{figure}
\centering
\includegraphics[width=0.49\textwidth]{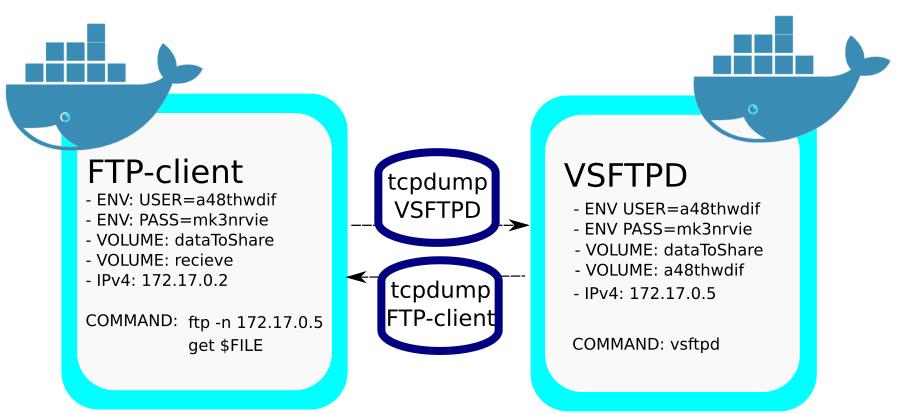}
\caption{Diagram of FTP scenario}
\end{figure}

We review the design of a prototypical capture scenario, namely, an FTP server and client interaction. The interaction is initiated by a single script, which allows the user to specify the length of the interaction, the number of times the interaction takes place as well as the specific subscenario. The script generates a random ftp username and password, creating the necessary \textit{User} directory on the host machine before calling the Docker-compose file which creates a bridge network. Subsequently, the necessary containers are then started which, in this case, consist of a VSFTPD server, a client with ftp instaled and two containers running tcpdump to capture all of the traffic emitted and received by the client and server respectively into separate \texttt{.pcap}-files. These \texttt{.pcap}-files are shared with the host machine via a shared volume. The host machine also shares:
 
\begin{itemize}
\item A \textit{dataToShare} volume containing files that can be downloaded by the client.
\item The \textit{User} directory with the server, which contains the same files as the \textit{dataToShare} folder.
\item An empty \textit{receive} folder with the client into which the files will be downloaded.
\item The random username and password is shared with the client container so it can authenticate itself to the server.
\end{itemize}
 
 Up to this point, no network traffic has been generated and the containers are now ready to begin communicating with one another. 
 For this particular interaction between an FTP server and client, we want to ensure that it is possible to capture the many ways in which an FTP session may develop. For instance, the client may seek to download files via the \texttt{get} command or the \texttt{put} command, alongside many other possibilities. We define 13 possible capture subscenarios intended to encapsulate a wide range of potential FTP sessions. These include downloading a single file using \texttt{get} or \texttt{put}, downloading every file using \texttt{mget} or \texttt{mput}, deleting every file from the server and requesting files from the server without the necessary authentication.

 After the scenario ends, both the \textit{User} directory and any downloaded files are removed from the host machine. The containers are then stopped and the bridge network is torn down. All necessary containers, volumes and scripts are in the same position prior to initiating the scenario --- barring any generated \texttt{.pcap}-files --- allowing for the scenarios to be started repeatedly with minimal human interaction. The \texttt{.pcap}-files are tagged with information about the time of creation, executed scenario and subscenario, and the container generating the traffic.

\subsection{Dataset creation}\label{Sec:datasetcreation}

Our framework generates network datasets consisting of a single interaction, but it is possible
to coalesce these datasets to create larger datasets with a wide variety of traffic, albeit with some caveats. Due to the networking constraints of the Docker virtual network, such as limitations regarding clashing ports, running many of our Docker scenarios simultaneously over a large period of time is unfeasible. Thus, to ensure that the generated traffic is suitably heterogeneous, numerous datasets must be generated before being coalesced into a main dataset. If done naively, this presents a problem. As discussed by Shiravi et al. \cite{shiravi2012toward}, merging distinct network data in an overlapping manner can introduce inconsistencies. For instance, if one wanted to create a dataset containing both normal webserver traffic and traffic originating from a Denial of Service attack, it would not work to generate these two datasets separately before merging them together. If these two events really did occur simultaneously, the high network throughput of the latter would likely effect the packet timings of the former. 

To avoid such inconsistencies, we create larger datasets by collecting data in consecutive chunks of fixed time.
Within each chunk, several scenarios are run simultaneously. All \texttt{.pcap}-files collected during a given chunk can be merged together. It is then simple to stitch together all of these chunks into a single \texttt{.pcap}-file using a combination of \texttt{Mergecap} \cite{mergecap} and \texttt{Editcap} \cite{editcap}. This allows us to shift the timings of each \texttt{.pcap}-file by a fixed amount such that all of our chunks occur in succession whilst maintaining the internal consistency of each chunk.

\subsection{Scenarios}\label{Sec:ExistScen}


Our framework contains 29 scenarios, each simulating a different benign or malicious interaction. The protocols underlying benign scenarios were chosen based on their prevalence in existing network traffic datasets.
These datasets consist of common internet protocols such as HTTP, SSL, DNS, and SSH. According to our evaluation, our scenarios can generate datasets containing the protocols that make up at least $87.8\%$ (MAWI), $98.3\%$ (CIC-IDS 2017), $65.6\%$ (UNSW NB15), and $94.5\%$ (ISCX Botnet) of network flows in the respective dataset.
Our evaluation shows that some protocols that make up a substantial amount of real-world traffic are glaringly omitted by current synthetic datasets, such as BitTorrent or video streaming protocols, which we decided to include.

In total, we produced 17 benign scenarios, each related to a specific protocol or application. Further scenarios can be added in the future, and we do not claim that the current list exhaustive. Most of these benign scenarios also contain many subscenarios where applicable.

The remaining 12 scenarios generate traffic caused by malicious behavior. These scenarios cover a wide variety of major attack classes including DoS, Botnet, Bruteforcing, Data Exfiltration, Web Attacks, Remote Code Execution, Stepping Stones, and Cryptojacking. 
Scenarios such as stepping stone behavior or Cryptojacking previously had no available datasets for study despite need from academic and industrial researchers.


We provide a complete list of implemented scenarios in Table \ref{tab:scen}.

\begin{table}
\begin{tabular}{l|l|r}
 \hline
 Name & Description & \#Ssc. \\
 \hline
 Ping & Client pinging DNS server & 1 \\
 Nginx & Client accessing Nginx server & 2\\
 Apache & Client accessing Apache server & 2\\
 SSH & Client communicating with & 5\\
 &SSHD server&\\
 VSFTPD & Client communicating with & 12\\
 &VSFTPD server&\\
 Scrapy & Client scraping website & 1 \\
 Wordpress & Client accessing Wordpress site & 1\\
 Syncthing& Clients synchronize files & 1\\
 &via Syncthing&\\
 mailx& Mailx instance sending & 2\\
 &emails over SMTP &\\
 IRC & Clients communicate via IRCd& 2\\
 BitTorrent & Download and seed torrents & 3 \\
 SQL & Apache with MySQL & 2\\
 NTP & NTP client & 2\\
 Mopidy & Music Streaming & 5\\
 RTMP & Video Streaming Server & 1\\
 WAN Wget & Download websites & 5 \\
 \hline
 SSH B.force & Bruteforcing a password & 3\\
 &over SSH&\\
 URL Fuzz & Bruteforcing URL & 1\\
 Basic B.force & Bruteforcing Basic & 2\\
 &Authentication&\\
 Goldeneye & DoS attack on Web Server & 1\\
 Slowhttptest & DoS attack on Web Server & 4 \\
 Mirai & Mirai botnet DDoS & 3\\
 Heartbleed & Heartbleed exploit & 1\\
 Ares & Backdoored Server & 3\\
 Cryptojacking & Cryptomining malware & 1\\
 XXE & External XML Entity & 3\\
 SQLi & SQL injection attack & 2 \\
 Stepstone & Relayed traffic using & 2\\
 &SSH-tunnels&\\
 \hline
\end{tabular}
\caption{Currently implemented traffic scenarios along with the number of implemented subscenarios}
\label{tab:scen}
\end{table}

\section{Validation experiments}\label{Sec:Experiments}

A framework that generates network traffic does not necessarily provide realistic and useful data. To evaluate the utility of our Docker framework, we construct a series of experiments. We have two goals in mind. First, we want to demonstrate that the traffic generated is sufficiently representative of real-world traffic.
Second, we want to demonstrate that having a framework to continually generate data compared to static datasets benefits evaluating the efficacy of intrusion detection systems.

The first experiment provides a general verification of the reproducability of our framework, which is required for guarantee the ground truth of the produced data. The second experiment demonstrates that the WAN-characteristics we emulate for our data make it quasi non-distinguishable from real WAN traffic. Our third experiment then demonstrates the advantage of unlimited data generation capabilities for training ML-based traffic classification.

\subsection{Reproducible scenarios}\label{Sec:deterministic}

To provide ground truth, we have to guarantee that our implemented scenarios and subscenarios are consistent and reproducible upon repeated execution. This applies both to consistency for external influences on the host, such as increased computational load, as well as internal consistency of the implemented script execution. 

\begin{figure}
\includegraphics[width=0.45\textwidth]{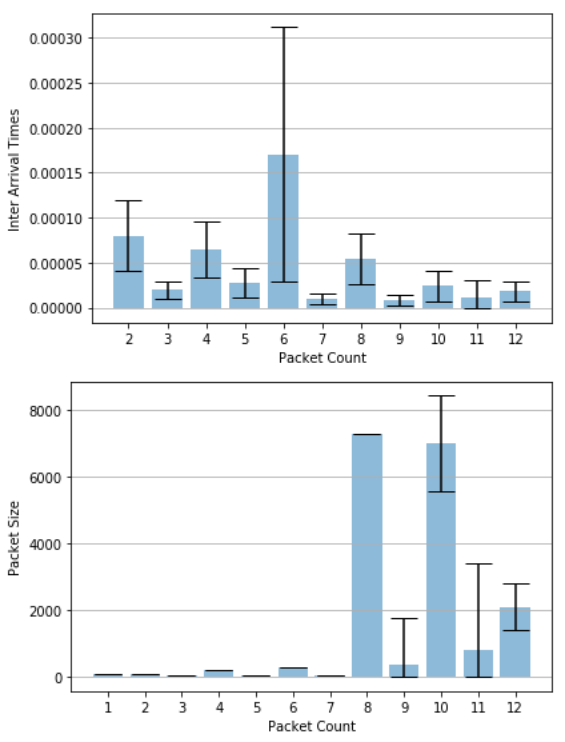} 
\caption{Means of IATs \& packet sizes along with standard deviation bars for the first twelve packets in the Apache scenario.}
\label{fig:size1}
\end{figure}

It is impossible to guarantee that each scenario will produce a truly `deterministic', or repeatable, output due to differences in network conditions, computational times, or input. Instead, we aim for our data to be \textit{reproducible up to networking and computational differences}. This means that when running a scenario multiple times, we expect the quantities of most packets to be largely identical. We do expect some packets to exhibit greater variation due to non-determinism in the underlying protocols, Fig. \ref{fig:size1} outlines this behavior in terms of interarrival times and packet sizes. 

To measure how consistent our scenarios are, we generate 500 \texttt{.pcap} files for three different implemented scenarios, namely the Apache, the VSFTPD, and the SSH scenario. These were generated consecutively under different host CPU load. We did not apply any delays or other NetEm traffic controls.

We assess the consistency of a scenario across different \texttt{.pcap} files by comparing all generated \texttt{.pcap} files pairwize. We measure this by the similarity of the connections captured. 

To test the similarity of two connections, we extract the sample distributions of the packet interarrival times and packet sizes overall, upstream and downstream. We define two connections as similar if the two distributions for each of these quantities pass an equality test. We use the two-sample Kolmogorov-Smirnov (K-S) test, a non-parametric statistical test for the equality of two continuous one-dimensional distributions \cite{massey1951kolmogorov}, with a p-value of $0.01$. 

As all tested files passed this similarity test, we conclude that these scenarios yield consistent and reproducible results. As other scenarios follow the same setup and launch commands, we expect the results to stay the same as long as the involved containers are consistent in their behavior.


\subsection{Explorating Artificial Delays}



Most traffic our framework generates is transported over Docker's virtual network and therefore does not succumb to problems associated with normal network congestion, such as packet loss, corruption and packets arriving out of order. A realistic dataset should include these phenomena, which is why we developed wrapping scripts that allow us to artificially add delays as well as packet loss and corruption, using NetEm. Choosing the parameters is not straightforward; it is not clear how close to real-world traffic such network emulation techniques are. This is especially true for packet delays, which are described by continuous distributions and often have temporal correlation.

Furthermore, the high effective bandwidth of the Docker virtual network resulted in traffic with extremely short inter-arrival times (IATs, defined as the time between two packet arrivals). Therefore, we devote considerable time to demonstrating that it is possible for traffic generated by our Docker framework to conform to real-world IAT distributions when altered using NetEm.


\subsubsection*{Datasets} We create two classes of datasets, one which is representative of `real-world' traffic, and one which has been generated from our Docker framework. For simplicity, we only consider datasets consisting of FTP traffic.

For the real-world dataset, we set up a containerized VSFTPD server running on a \emph{Google Compute} virtual machine located in the Eastern United States, and a containerized FTP client on our local host. We then ran a series of our scripted interactions between the two machines, generating 834 megabytes of data in 250964 packets. These interactions consisted of several FTP commands with various network footprints. We collect all data transmitted on both the server and the client. We call this data the \textit{Non-Local} dataset.

We then repeat this process using the same container setup, but across the Docker virtual network on a local machine. We repeat this process several times, generating several \textit{Local} datasets under a variety of emulated network conditions, discussed in Section \ref{Sec:method1}. Our \textit{Local} datasets vary slightly in size, but are all roughly 800 megabytes with 245000 packets. 


\subsubsection*{Methodology}
\label{Sec:method1}

NetEm allows us to introduce packet delays according to a variety of distributions, namely uniform, normal, Pareto and Paretonormal\footnote{This Paretonormal distribution is defined by the random variable $Z = 0.25*X + 0.75*Y$, where $X$ is a random variable drawn from a normal distribution and $Y$ is a random variable drawn from a Pareto distribution.}. Furthermore, NetEm adds delays according to modifiable distribution tables, and so it is trivial for us to add a Weibull distribution, which along with Pareto distributions have been shown to closely model packet IATs \cite{arfeen2013role,paxson1995wide}.
In total, we test the efficacy of four distributions to model inter-arrival times --- normal, Pareto, Paretonormal and Weibull. 
 
We generate several \textit{Local} datasets by delaying traffic according these distributions, performing an exhaustive grid search over their means and standard deviations. Initial experiments revealed that introducing delays with a mean in the range of 40 ms to 70 ms produced the best results. Setting the jitter of the distribution too high resulted in the repeated arrival of packets out of order, therefore we further limit the grid search to jitter values in 5ms intervals up to half of the value of the mean. In total, we generate 88 \textit{Local} datasets.

Our goal is to discover the \textit{Local} dataset whose packet timings most closely resemble those of our \textit{Non-Local} dataset. To do this, we extract the IATs and packet sizes from our datasets on a packet-by-packet basis and store these results in arrays. We measure the similarity between two of these arrays by training a Random Forest classifier to distinguish between them. We say that if the Random Forest correctly classifies each packet with a success rate of only 50\% then it is no better than randomly guessing and, as such, the inter-arrival times of these two arrays are indistinguishable from one another for the Random Forest.

To perform this measurement, we concatenate one \textit{Local} dataset array with our \textit{Non-Local} dataset array, label the entries and then shuffle the rows. We proportion this data into a training set and a testing set using an 80-20 split. We then feed this training data into a Random Forest with 1000 trees and fixed seed, and then record the accuracy of this Random Forest on the test set. We repeat this process for every single \textit{Local} dataset.

\subsubsection*{Results}

\begin{figure}
\captionsetup{justification=centering}
\centering
\includegraphics[width=0.45\textwidth]{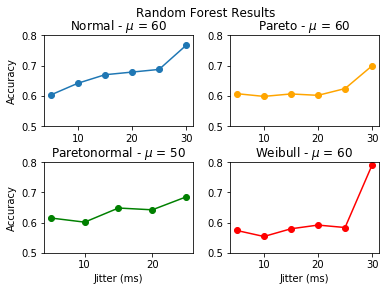}
\caption{Results of Random Forest Classifier for a given distribution at the best performing delay mean $\mu$. Note that a score of .5 indicates total indistinguishability.}
\label{Fig:rf_graph}
\end{figure}

Table \ref{tab:results-iat_rf} summarizes the values of the mean and jitter for a given distribution that produced the worst results from the random forest classifier.

\begin{table}[ht!]
\begin{center}
\begin{small}
\begin{sc}
\begin{tabular}{ccccc}
\hline
Distribution & Mean & Jitter & RF Accuracy\\
\hline
No Delays (Baseline) & 0 & 0ms & 0.8176 \\
Constant Delay & 40ms & 0ms & 0.6730 \\
Normal & 60ms & 5ms & 0.6028 \\
Pareto & 60ms & 10ms & 0.5979 \\
Paretonormal & 50ms & 10ms & 0.6015 \\
Weibull & 60ms & 10ms & 0.5540 \\
\hline
\end{tabular}
\end{sc}
\end{small}
\caption{Worst Random Forest accuracy rates for a given distribution}
\label{tab:results-iat_rf}
\end{center}
\vskip -4mm
\end{table}

To establish a baseline, we compare the traffic generated from our Docker scenario to that of the \textit{Google Compute} data with no added delays. In this case, the Random Forest was able to distinguish between the two datasets, achieving an accuracy of over 90\%. The classification accuracy is worsened considerably by introducing network delays, with the best results being achieved using a Weibull distribution with a mean of 60 ms and a jitter of 10 ms, leading to an accuracy of just $55\%$. Results for Pareto and Weibull distributions seem to yield consistent results for differing jitter values.
Although not completely indistinguishable, this proves that using NetEm we can emulate WAN properties very closely.

\subsection{Advantages of Dynamic Dataset Generation}

Having examined whether our Docker framework is capable of emulating real-world IATs, we explore their utility in traffic classification to demonstrate the advantages that our framework provides compared to static, unlabeled datasets.

Machine-learning techniques are a popular tool for traffic classification, with many successful published classifiers. Furthermore, inter-packet arrival times have been shown to be a discriminative feature \cite{zander2005automated,nguyen2008survey}. However, these methods considered datasets consisting of completed traffic flows, limiting their use in, say, a stateful packet inspector. On-the-fly classifiers are also successful. Jaber et al. \cite{jaber2011can} showed that a K-means classifier can classify flows in real-time solely based on IATs with precision exceeding 90\% for most protocols within 18 packets. Similarly, Bernaille et al. \cite{bernaille2006traffic} demonstrated that a K-means classifier can precisely classify traffic within five packets using only packet size as a feature. 

 However, Jaber et al. \cite{jaber2011can} only evaluated their traffic classifier with training and testing data drawn from the same dataset containing traces of a single network; there is no measure of how this model may generalize to other networks with differing conditions. Furthermore, they were limited to using unsupervised machine learning algorithms to classify their traffic as their datasets had no ground truth. 

We attempt to replicate these results within our Docker framework with some adjustments. As we can generate a fully accurate ground truth, we attempt to segregate application flows based on their packet IATs using supervised learning techniques. Moreover, we then measure this model's ability to generalize by expanding our dataset to include traffic from networks with differing bandwidth and latency.

\subsubsection*{Data \& Preprocessing}

Our goal is to measure a classifier's ability to generalize across datasets. Therefore we construct two datasets using our Docker framework, both containing the same number of network traces from the same containers. 

For our first dataset, we generate \texttt{.pcap}-files, each containing traffic from one of 16 different classes: HTTP (Client \& Server), HTTPS (Client \& Server), RTMP (Client, Server \& Viewer), SSH (Client \& Server), FTP (Client \& Server), IRC (Client \& Server), SMTP, SQLi and DoS traffic. To prevent class imbalance, we generate 200 \texttt{.pcap}-files for each of the 16 classes, resulting in 3200 total files. To more accurately emulate potential network conditions, we use our NetEm scripts to apply a unique delay to every container involved in a scenario. These delays follow a Pareto distribution with random mean between 0 and 100 milliseconds and random jitter between 0 and 10 milliseconds. We then preprocess this data by removing all but the first 12 packets of each \texttt{.pcap}-file. We extract the 11 inter-arrival times separating the 12 packets, which act as our feature vectors. We collect these feature vectors for each class along with a class label, and store collected feature vectors from all 3200 \texttt{.pcap}-files in a 12 x 3200 array. We call this our \textit{Primary} dataset.

We then repeat this process to generate a second dataset, changing the properties of our emulated network. Again, we delay all traffic using a Pareto distribution, however, this time we select a random mean in the range of 100 to 500 milliseconds and random jitter between 0 and 50 milliseconds. The subsequent preprocessing of our data remains unchanged. We call this our \textit{Secondary} dataset.

\subsubsection*{Methodology}

\label{Sec:exp2_method}
First, we attempt to reproduce the results presented by Jaber et al. \cite{jaber2011can} by training a Random Forest with 100 trees to classify application flows based on packet IATs. We do this by proportioning our Primary dataset into training and testing sets using an 80-20 split. We then train and test our Random Forest repeatedly, first considering the classification accuracy based on the IATs of only the first two packets, then the first three packets and so on, up to 12 packets. We record the resulting confusion matrix for each round and calculate the precision and recall rates of our classifier.

Having trained the classifier, we measure its ability to generalize by repeating the above experiment, but replacing the test set with the Secondary dataset.

\subsubsection*{Results}

\begin{figure}

\subfloat{%
 \includegraphics[width=0.4\textwidth]{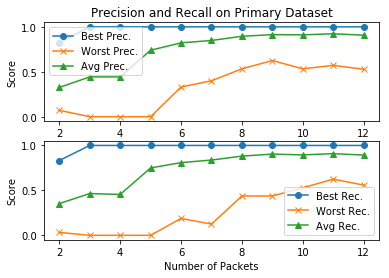}
}

\subfloat{
 \includegraphics[width=0.4\textwidth]{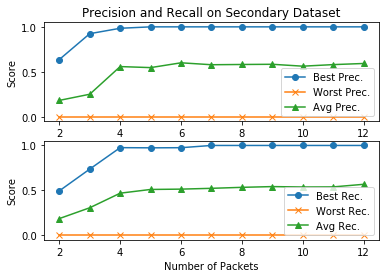}
}

\caption{Results of Random Forest Classification on Primary dataset (Above) and Secondary dataset (Below)}
\label{Fig:Primary}
\end{figure}

 After each run of our Random Forest on our Primary dataset, we gather the True Positive ($T_{P}$), False Positive ($F_{P}$) and False Negative ($F_{N})$ rate for each class. We then calculate their precision, defined as $ \frac{T_P}{T_P + F_P}$, and recall, defined as $ \frac{T_P}{T_P + F_N}$, values. in Fig. \ref{Fig:Primary}, we see that our average precision and recall across the classes exceeds 0.9 after 10 IATs. Furthermore, after 12 packets our DoS and SQLi data is classified with precision and recall rates of 1.0 and 1.0 and 0.9462 and 0.9322 respectively.

These results does not hold when we test the classifyer on our Secondary dataset. As seen in Fig. \ref{Fig:Primary}, we see a substantial decrease in our average precision and recall rates, achieving a maximum of 0.5923 and 0.5676 respectively. Moreover, after four packets, increasing the number of IATs in our dataset provides little additional benefit. Although some services generalized well, such as IRC-client and IRC-server, others failed to be classified, with every single SMTP feature being classified as HTTP-client. We also see a substantial drop-off in the classification of malicious traffic, with the precision rates of DoS and SQLi data not exceeding 0.6.

These diverging results demonstrate the necessity of dynamic dataset generation for evaluation purposes. Researchers evaluating their methods only on a dataset with fixed properties such as the Primary dataset might receive overoptimistic results. The capability of generating two or more datasets with the same traffic classes, but otherwise differing properties, provides a more realistic evaluation.

\section{Conclusions}\label{Sec:Conclusion}

In this paper, we outlined four requirements a modern dataset has to fulfil to strengthen the training of intrusion detection systems. We then proposed a Docker framework capable of generating network intrusion datasets that satisfy these conditions. The major design advantage of this framework are the isolation of traffic scenarios into separate container arrangements, which allows the extension of new scenarios and detailed implementation of subscenarios as well as the capture of ground truth of the computational origins of individual traffic events. Furthermore, containerization enables the generation of traffic data at scale due to containers being light-weight and easily clonable.

We verified the realism of the generated traffic and the corresponding ground truth information with two experiments, and demonstrated the usefulness of the framework in another experiment.
Presently, our framework consists of 29 scenarios capable of producing benign and malicious network traffic. Several of these scenarios, such as the \emph{BitTorrent} or the \emph{Stepping-Stone} scenario, provide novel traffic data of protocols or behaviours that has not been widely available to researchers previously.

\subsection{Difficulties and limitations}

Our framework is building network traffic datasets from a small-scale level up by coalescing traffic from different fine-grained scenarios together. While this provides great insight into small-scale traffic structures, our framework will not replicate realistic network-wide temporal structures, such as port usage distributions or long-term temporal activity. These quantities would have to be statistically estimated from other real-world traffic beforehand to allow our framework to emulate such behavior reliably. Other datasets such as UGR-16 use this approach to fuse real-world and synthetic traffic and are currently better suited to build models of large-scale traffic structures.

Working with Docker containers can sometimes complicate the implementation of individual scenarios compared to working with VMs. Although several applications are officially maintained Docker containers that are free from major errors, many do not. For instance, in the \textit{BitTorrent} scenario, most common command line tools, such as \texttt{mktorrent}, \texttt{ctorrent} and \texttt{buildtorrent}, failed to actually produce functioning torrent files from within a container due to Docker's union filesystem. Furthermore, due to the unique way in which we are using these software packages, unusual configuration settings are sometimes needed. 

Lastly, capturing \texttt{.pcap}-files from each container can quickly exceed available disc space when generating traffic at scale. Depending on specific research requirements, it is advizable to add filtering or feature extraction commands to the scenario execution scripts to enable traffic preprocessing in real-time.

\subsection{Future work}

Our traffic generation framework is designed to be expandable and there are many avenues for future work. The continual development of scenarios and subscenarios would improve the potential realism of datasets generated using the framework. The addition of more malicious scenarios would enable a more detailed model evaluation and improve detection rate estimation. 
Another future improvement for framework is to add scripts that emulate the usage activity of individual scenarios by a user or a network.

Although ground truth for particular traffic traces is provided by capturing \texttt{.pcap}-files for each container individually, we have not implemented a labelling mechanism yet for the dataset coalescence process. Though not technically difficult, some thought will have to be put how such labels would look like to satisfy different research demands.
Furthermore, the Docker platform provides the functionality to collect system logs via the \texttt{syslog logging driver}. We plan on implementing their collection in the future, where they could act either as traffic labels providing more ground truth details, or act as a separate data source that complements the collected traffic.

We wish to publish this framework to a wider audience, allowing for further modification. This will be done using a GitHub repository, which contains both the implemented capture scenarios as well as the corresponding container images.



\section{Acknowledgments}

We are grateful to  British Telecommunications PLC who are supporting the PhD research of the first
author in the UK EPSRC CASE scheme, giving invaluable guidance on the
needs and possibilities of intelligent security tools and their
evaluation.  Part of this paper draws on the 2019 MSc dissertation of
the second author.  
Nikola Pavlov helped with some of the earlier implementation.  The
third author was supported by The Alan Turing Institute under the
EPSRC grant EP/N510129/1 and the Office of Naval Research ONR NICOP
award N62909-17-1-2065.



\bibliographystyle{ACM-Reference-Format}
 
\bibliography{ACSAC_DYNAMICS}

\end{document}